\documentclass[aps,prl,twocolumn,twoside,floatfix]{revtex4}

\usepackage{graphicx}
\usepackage{color}

\begin{document}
\global\long\def\Al{Al$^{+}$ }
 \global\long\def\Alns{Al$^{+}$}
 \global\long\def\AlTS{$^{27}$Al$^{+}$ }
 \global\long\def\AlTSns{$^{27}$Al$^{+}$}
 \global\long\def\Be{Be$^{+}$ }
 \global\long\def\Bens{Be$^{+}$}
 \global\long\def\BeN{$^{9}$Be$^{+}$ }
 \global\long\def\BeNns{$^{9}$Be$^{+}$}
 \global\long\def\Mg{Mg$^{+}$ }
 \global\long\def\Mgns{Mg$^{+}$}
 \global\long\def\MgTF{$^{25}$Mg$^{+}$ }
 \global\long\def\MgTFns{$^{25}$Mg$^{+}$}
 \global\long\def\Hg{Hg$^{+}$ }
 \global\long\def\clocktransition{$^{1}$S$_{0}$$\leftrightarrow$$^{3}$P$_{0}$ }
 \global\long\def\clocktransitionAlmF#1#2{$|^{1}S_{0},m_{F}=#1\rangle$$\leftrightarrow$$|^{3}P_{0},m_{F}=#2\rangle$}
 \global\long\def\diffAlAl{$-1.8\times10^{-17}$ }
 \global\long\def\diffpmAlAl{$(-1.8\pm0.7)\times10^{-17}$ }
 \global\long\def\diffpmAlAlns{$(-1.8\pm0.7)\times10^{-17}$}
 \global\long\def\diffAlAlns{$-1.8\times10^{-17}$}
 \global\long\def\sigmaAlAl{$2.5\times10^{-17}$ }
 \global\long\def\uncStat{$7.0\times10^{-18}$ }
 \global\long\def\uncStatns{$7.0\times10^{-18}$}
 \global\long\def\systAlMg{$8.6\times10^{-18}$ }
 \global\long\def\systAlMgns{$8.6\times10^{-18}$}
 \global\long\def\stabAlMg{$2.8\times10^{-15}$}
 \global\long\def\systAlBe{$2.3\times10^{-17}$ }
 \global\long\def\systSr{$1.5\times10^{-16}$ }
 \global\long\def\systHg{$1.9\times10^{-17}$ }
 \global\long\def\dnn{\Delta\nu/\nu}
 \global\long\def\xfertransition{$^{1}$S$_{0}$ $\rightarrow$ $^{3}$P$_{1}$ }
 \global\long\def\MMns{EMM}
 \global\long\def\MM{EMM }

\newcommand{\dd}{|\!\!\downarrow_1\downarrow_2\rangle}
\newcommand{\du}{|\!\!\downarrow_1\uparrow_2\rangle}
\newcommand{\ud}{|\!\!\uparrow_1\downarrow_2\rangle}
\newcommand{\uu}{|\!\!\uparrow_1\uparrow_2\rangle}
\newcommand{\da}{|\!\!\downarrow\rangle}
\newcommand{\ua}{|\!\!\uparrow\rangle}

\title{Quantum coherence between two atoms  beyond $Q=10^{15}$}

\author{C. W. Chou}

\email[]{chinwen@nist.gov}

\author{D. B. Hume}

\author{M. J. Thorpe}

\author{D. J. Wineland}

\author{T. Rosenband}

\affiliation{Time and Frequency Division, National Institute of Standards and
Technology, Boulder, Colorado 80305}

\date{\today}
\begin{abstract}
We place two atoms in quantum superposition states and observe coherent phase evolution for $3.4\times10^{15}$ cycles.
Correlation signals from the two atoms yield information about their
relative phase even after the probe radiation has decohered.
This technique was applied to a frequency comparison of two \AlTS ions,
where a fractional uncertainty of $3.7^{+1.0}_{-0.8}\times10^{-16}/\sqrt{\tau/s}$  was observed. Two measures of the Q-factor are reported: The
Q-factor derived from quantum coherence is $3.4^{+2.4}_{-1.1}\times10^{16}$, and the spectroscopic
Q-factor for a Ramsey time of 3 s is $6.7\times10^{15}$. As part of this experiment, we demonstrate a method to detect the individual quantum states of
two \Al ions in a \Mgns-\Alns-\Al linear ion chain without spatially resolving the ions.
\end{abstract}
\maketitle

Coherent evolution of quantum superpositions follows directly from Schr\"odinger's
equation, and is a hallmark of quantum mechanics. Quantum systems with a high degree of coherence are desirable for sensitive measurements and for studies in quantum control. Typically, quantum superposition states quickly decohere due to uncontrolled interactions between the system and its environment. However, through careful isolation and system preparation, quantum coherence
has been observed in naturally occurring systems
 including photons and atoms, as well as in engineered macroscopic systems
\cite{haroche06,Nature2008QCoherence,leggett02,OConnell2010}.
 In order to observe the coherence time of a system, it must be compared to a reference system that is
at least as coherent, a requirement that can be difficult to satisfy, particularly in systems with the highest degree of coherence.
In atomic physics, quality (Q-) factors as high as $1\times10^{14}$ to $4\times10^{14}$ \cite{Rafac2000Hg, Boyd2006Coherence, Chou2010Relativity} have been observed with laser spectroscopy, where the linewidths are often limited by laser noise rather than atomic decoherence. In this report we apply a recent spectroscopic technique~\cite{Chwalla2007Corr} to directly observe atomic coherence beyond the laser limit and probe an atomic resonance with a Q-factor above $10^{15}$.

Historically, M\"{o}ssbauer spectroscopy with $\gamma$-rays has exhibited
the highest relative coherence, as quantified by the spectroscopic
Q-factor (the ratio of oscillation frequency to observed resonance linewidth).
Values as high as $8.3\times10^{14}$ are observed~\cite{Potzel1976} in the 93.3~keV radioactive decay of $^{67}$Zn, limited by the nuclear lifetime of 13.4~$\mu$s. In those measurements, two separate crystals that contained $^{67}$Zn nuclei were compared. One sample provided the probe radiation, while the other
served as the resonant absorber. The M\"ossbauer
method might be extended to characterize optical transitions in atoms~\cite{Dehmelt1988HQ},
but here we use a method based on Ramsey spectroscopy
in which the phase fluctuations of the probe source are rejected as common-mode noise \cite{Chwalla2007Corr}, enabling Ramsey times much longer than the probe coherence time. Other experiments that compare
pairs of microwave \cite{Bize2000a} or optical clocks~\cite{Katori2010ICAP} use a related technique to reduce the Dick-effect noise~\cite{Dick1987DickEffect,Lodewyck2010Dick} that can limit the stability of frequency comparisons.

\begin{figure}
\includegraphics[width=0.45\textwidth]{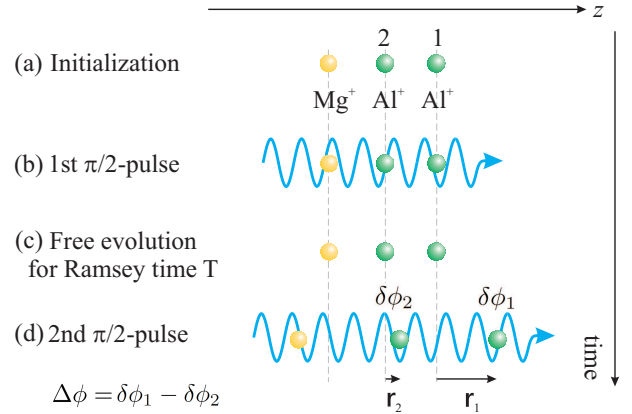} \caption[The protocol]{(Color online) Illustration of the protocol.
(a) The detected states from the previous Ramsey experiment
serve as the initial states for the the current measurement. (b) The first $\pi/2$-pulse is applied. This is accomplished by a laser
beam whose axis coincides with the axis of the ion array. (c) The
clock state superpositions freely evolve. (d) The
spacing is adjusted at the end of the free-evolution period to vary
the differential phase $\Delta\phi$. This is followed immediately by the second $\pi/2$-pulse. At the end of the sequence,
the final states are detected to obtain the correlation. }
\label{Fig1}
\end{figure}

In the experiment reported here, atomic superposition states evolve coherently for up to 5 s at a frequency of $1.12\times10^{15}$ Hz. Following Chwalla \textit{et al}.~\cite{Chwalla2007Corr}, a Ramsey
pulse sequence \cite{Ramsey1956} is simultaneously applied to two trapped $^{27}$Al$^+$ ions, labeled $i\in\{1,2\}$ (see Fig. 1). The probe radiation for both ions is derived from the same source. Each ion is initialized in one of the two quantum states that make up the clock transition (clock states), which need not be the same for both ions. Immediately prior to the second $\pi/2$-pulse, a variable displacement $\mathbf{r_i}$ is applied to the ions. This Ramsey sequence induces a state change with probability $p_{i}=(1+\cos{\delta\phi_{i}})/2$, where $\delta\phi_{i}=\phi_{L} + \mathbf{k} \cdot \mathbf{r_i} - \phi_i$
is the difference between the phase accumulated by the laser ($\phi_{L}$+$\mathbf{k} \cdot \mathbf{r_i}$) and ion ($\phi_i$) during the free-evolution period $T$, and $\mathbf{k}$ is the laser beam wavevector, $\mathbf{k}=\hat{z}2\pi/(267\text{ nm})$.
The correlation probability (the probability that both ions make a
transition, or both do not make a transition) is then $P=[2+\cos{(\delta\phi_{1}-\delta\phi_{2})}+\cos{(\delta\phi_{1}+\delta\phi_{2})}]/4$.
Here the relative phase, $\delta\phi_{1}-\delta\phi_{2}$, is independent of $\phi_{L}$, which is uniformly randomized over the interval $[0,2\pi)$ with a pseudo-random number generator. Without knowledge of $\phi_{L}$, the probability
of correlated transitions is
\begin{equation}
P_c=\frac{1}{2\pi}\int^{2\pi}_{0}Pd\phi_L=\frac{1}{2}+\frac{C}{2}\cos{\Delta\phi},\label{Pc}
\end{equation}
where $\Delta\phi=\phi_{2}-\phi_{1}+\mathbf{k}\cdot(\mathbf{r_1-r_2})$ and $C\equiv P_{c,\text{ max}}-P_{c,\text{ min}}\leq\frac{1}{2}$ is the contrast.

The correlation signal $P_c$ provides a measurement of the differential phase evolution of the two \Al ``clock'' ions similar to the measurement of differential phase between source and absorber in M\"{o}ssbauer spectroscopy. Its statistical properties are equivalent to that of a single-ion Ramsey experiment with reduced contrast, and the ultimate measurement uncertainty is determined by quantum projection noise~\cite{WMI1993ProjectionNoise}. When $|\Delta\phi|$ is kept near $\pi/2$, the statistical uncertainty of the ion-ion fractional frequency difference, or measurement instability, is $\sigma(\tau)\equiv\sigma_\nu/\nu=(2\pi\nu C \sqrt{T\tau})^{-1}$, where $\tau$ is the total measurement duration, $\sigma_\nu$ is the uncertainty in the measured frequency difference $(\phi_2-\phi_1)/(2\pi T)$, and $\nu\approx1.12$ PHz is the transition frequency. Importantly, the free-evolution period $T$ is not limited by laser phase noise.

In the experiment, a linear Paul trap confines one \MgTF ion and two \Al ions in an array~\cite{CWC2010AlAl,DBH2007detection} along the trap z-axis (Fig.~\ref{Fig1}).
The motional frequencies of a single \Mg in the trap
are $\{f_{x},f_{y},f_{z}\}=\{5.13,6.86,3.00\}$~MHz. The ions are
maintained in the order of \Mgns-\Alns-\Al (inter-ion spacing 2.69 $\mu$m) by periodically adjusting the
trap conditions and verifying via \Mg spectroscopy the frequency of the ``stretch'' mode of motion, whose value is 5.1 MHz for the correct order~\cite{DBH2010Thesis}.

The two states involved in the \Al clock transition, $\da \equiv |^{1}S_{0}$, $m_{F}=5/2\rangle$ and $\ua \equiv |^{3}P_{0}$, $m_{F}=5/2\rangle$, are detected with an adaptive quantum non-demolition
process \cite{DBH2007detection}.  The present implementation
distinguishes all four states $\dd$, $\du$, $\ud$, and $\uu$ by observing \Mg fluorescence after
controlled interactions between the \Al and \Mg ions. Individual state detection relies on the two \Al ions having different
amplitudes in several motional eigenmodes, which affects the state-mapping probability onto the \Mg ion. Information from several
measurements is combined in a Bayesian process~\cite{DBH2007detection}, to determine the most
likely state of the two \Al ions with typically 99~\% probability in an
average of 30 detection cycles (approximately 50 ms total duration).
  We note that this technique
allows individual state detection of two ions in the same trap,
without the need for high spatial-resolution optics.

\begin{figure}
\vspace{-0.5 cm}
\includegraphics[width=0.55\textwidth]{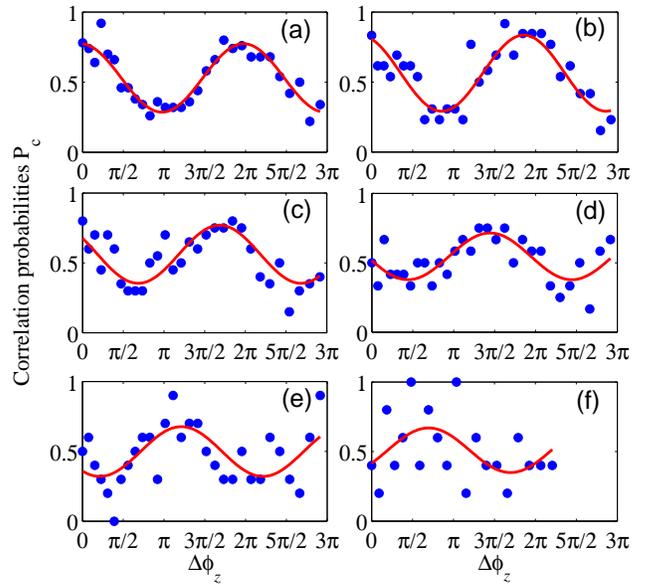} \vspace{-1.5 cm}\caption[Ramsey fringes]{(Color online) Correlation probabilities $P_c$ versus $\Delta\phi_z$ for various Ramsey times: (a) 0.1~s, 1500 probes; (b) 0.5~s, 600 probes;(c) 1~s, 600 probes; (d) 2~s, 360 probes; (e) 3~s, 300 probes; (f) 5~s, 100 probes. Dots: measurement outcomes; lines: maximum-likelihood fits to the fringes.}
\label{Fig2}

\end{figure}

The Ramsey experiments use $\pi/2$-pulse durations of 1.2~ms and are carried out for various free-evolution periods $T$. For each $T$, $\Delta\phi_z\equiv\mathbf{k}\cdot (\mathbf{r_{1}-r_{2}})$ is varied from
0 to beyond $2\pi$ to characterize the correlation. The duration required to shift the positions by $\mathbf{r_{i}}$ is approximately 10~ms. Figure~\ref{Fig2} shows
the correlation signals for $T$ between 0.1 and 5 s. Currently, collisions between the ions and background gas make it impractical to generate sufficient statistics for $T$ greater than 5 s. The collisions result in changes of ion order and loss of ions due to chemical reactions.

\begin{figure}
\vspace{-0.5 cm}
\includegraphics[width=0.45\textwidth]{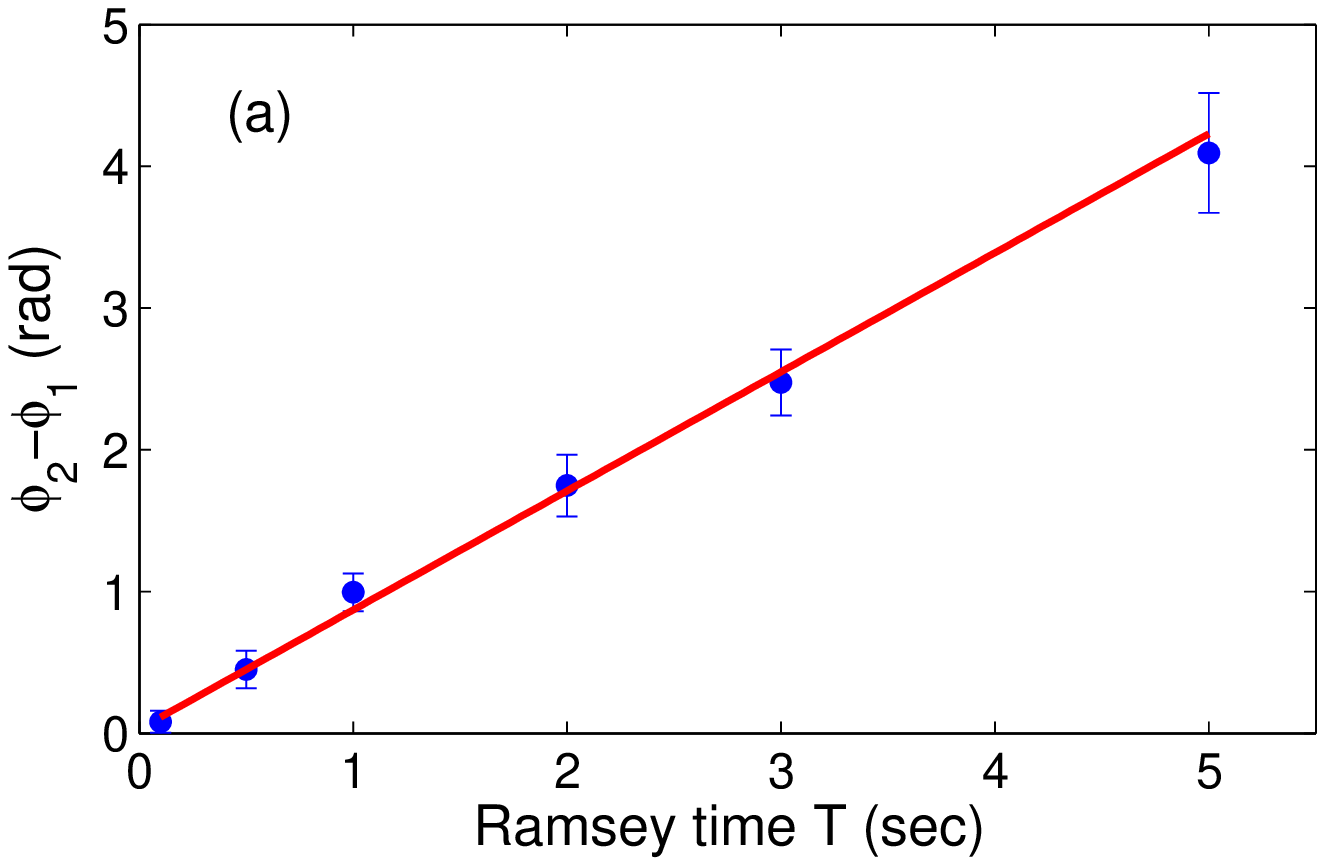}
\includegraphics[width=0.45\textwidth]{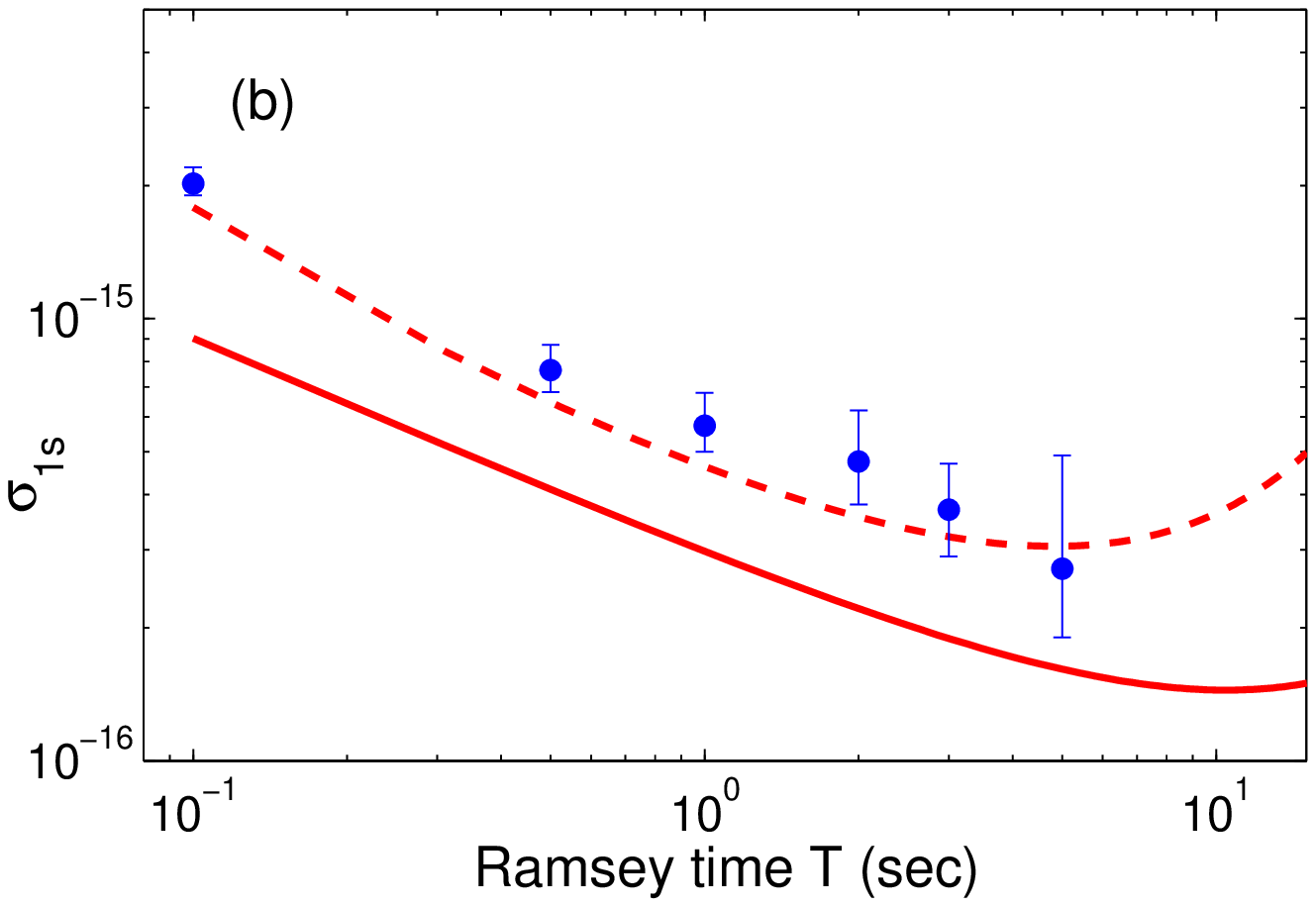}
\caption[contrast and phase]{(Color online) (a) Differential phase $\phi_2-\phi_1$ versus Ramsey time $T$. The solid line is a linear fit, with
slope $0.84\pm0.06$ rad/s. (b) Measurement uncertainty extrapolated to 1 s averaging time as a function of Ramsey time. Dots: measurement results, where the uncertainties are derived from the uncertainties in the contrast $C$; solid line: theoretical lifetime-limited instability, where only phases corresponding to $\Delta\phi\approx\pm\pi/2$ are probed; dashed line: expected experimental instability, with $\Delta\phi$ uniformly distributed over $[0,2\pi)$. The dashed line is derived from the measured coherence time of 9.7~s, and an approximate overhead of 100 ms per Ramsey measurement, which reduces the duty cycle.}
\label{Fig3}
\vspace{-0.2 cm}
\end{figure}

The phase difference $\phi_2-\phi_1$, and thus the frequency difference, between the two \Al ions can be determined from the phases of the $P_c$ fringes in Fig.~\ref{Fig2}. In the experiment, we apply a magnetic field gradient of $dB/dz=1.32\pm0.33$
mT/m, as measured by monitoring the frequency of the $|F=3\text{, }m_{F}=-3\rangle\rightarrow|F=2\text{, }m_{F}=-2\rangle$
magnetic-field dependent transition in the \MgTF $3s\text{ }S_{1/2}$ ground
state hyperfine manifold, when the \Mg position along the trap axis
is adjusted. This gradient induces a fractional frequency shift $(\nu_2-\nu_1)/\nu=1.32\pm 0.33 \times10^{-16}$
between the $\da\leftrightarrow\ua$ transitions of the two \Al ions. The phases of the $P_c$ fringes, determined by maximum-likelihood fits~\cite{Sivia1996},
increase linearly with $T$, as shown in Fig.~\ref{Fig3}a. A linear
fit has a slope of $0.84\pm0.06$ rad/s, corresponding to a measured shift of $1.19\pm0.08\times10^{-16}$,
 in agreement with the shift caused by the magnetic-field gradient. All reported uncertainties represent a 68~\% confidence interval.

We derive the contrast $C$ from the maximum-likelihood fits to the data in Fig.~\ref{Fig2}. An exponential fit of $C$ versus $T$ yields a relative coherence time $T_C$ of $9.7^{+6.9}_{-3.1}$~s, corresponding to a Q-factor ($Q=\pi \nu T_C$~\cite{Vion2002Q}) of $3.4^{+2.4}_{-1.1}\times10^{16}$. A uniform prior distribution of $T_C$ on the interval 0~s to 25~s is assumed. The measured coherence time is compatible with the expected result, which is given by the lifetime $T'=20.6\pm1.4$~s~\cite{TR2007Al3P0observed} of the \Al $|^{3}P_{0}\rangle$ state.  When viewed in terms of Ramsey spectroscopy, for $T=3$~s, the full-width-at-half-maximum of the Ramsey signal corresponds to a Q-factor of $2\nu T=6.7\times10^{15}$.

 The current protocol could significantly reduce the total duration of future high-precision measurements with atomic clocks.  Figure~\ref{Fig3}b shows the measurement uncertainties extrapolated to 1~s ($\sigma_{1s}$) versus Ramsey time $T$. The long-term statistical uncertainty is then $\sigma(\tau)=\sigma_{1s}/\sqrt{\tau/s}$ for a measurement duration $\tau$. Note that, for $T=3$~s, the frequency difference between the two \Al ions
can be determined with a fractional uncertainty $\sigma=1.1\times10^{-17}$
in a 1126 s measurement (900 s integrated free-evolution time), which can be extrapolated
to infer a relative measurement uncertainty $\sigma_{1s}=3.7\times10^{-16}$. This result may be compared to a recent frequency difference measurement of two \Al clocks, where 65,000 s were required to reach the same uncertainty of $1.1\times10^{-17}$~\cite{Chou2010AlAl}.
In general, the lifetime-limited contrast is $C=\frac{1}{2}\exp(-T/T')$, yielding an instability of $\sigma(\tau)=\exp(T/T')/(\pi\nu\sqrt{T\tau})$, which is shown for \Al in Fig.~\ref{Fig3}b (solid line). The optimal probe time of $T=T'/2$ yields $\sigma_{1s}=1.4\times10^{-16}$.

\begin{figure}
\includegraphics[width=0.32\textwidth]{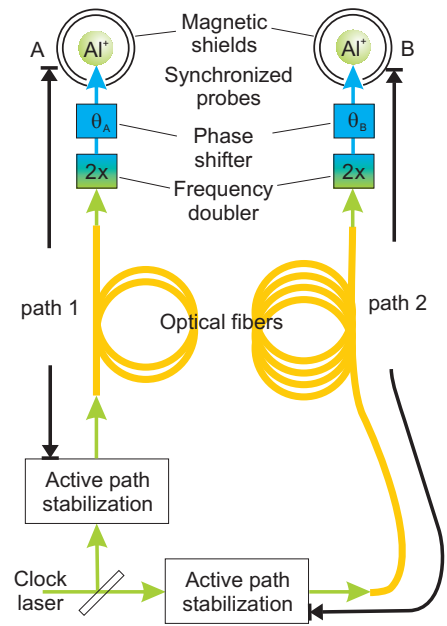} \caption[Clock comparison]{(Color online) Proposed frequency cmparison of remote optical clocks, here
based on \Al ions. The paths 1 and 2 that direct the clock laser light
to the ions need to be controlled so that they can faithfully transmit
the light without introducing additional phase noise. Local frequency
fluctuations, such as those caused by fluctuating magnetic fields,
should be minimized. The free evolution periods need to be synchronized
so that the atoms are subjected to the same phase noise in the Ramsey pulses,
the effect of which cancels in the protocol.}
\label{Fig4}
\vspace{-0.2 cm}
\end{figure}

  Although we have used the technique to measure two ions in the same trap, it may also be applied to clocks at different
locations. A proposed frequency comparison of remote optical
clocks is depicted in Fig.~\ref{Fig4}. Note, however, that due to the requirement that $\phi_L$ be the same for both clocks, this technique is limited to comparisons between clocks operating at similar frequencies. Although the clocks need not be identical, the differential phase, $\Delta\phi$, must be known well enough to make phase errors of $\pi$ unlikely. In order to retain control
over the differential phase, the individual paths (paths
1 and 2 in Fig.~\ref{Fig4})  need to be phase-stabilized and the Ramsey pulses at the two locations need to
be synchronized so that the two clocks experience
the same laser phase noise. For ions with very long radiative lifetimes, the same technique could be used to compare two ion samples, each composed of maximally entangled states~\cite{Leibfried2005Cat,Monz201014ionsEnt}.

 A similar approach can be taken in comparisons of two clocks composed of many unentangled atoms.  The measurement protocol is again based on synchronized Ramsey pulses where the free-evolution time $T$ exceeds the laser coherence time.  The two clocks (labeled $X\in\{A,B\}$) measure transition probabilities $p_{X}=\frac{1}{2}[1+\cos{(\phi_{X}-\phi_L-\theta_{X})}]$, and the quantity of interest $\delta\phi_{AB}=\phi_A-\phi_B$ is determined from $\delta\phi_{AB}=\cos^{-1}{(2p_A-1)}-\cos^{-1}{(2p_B-1)}+\theta_A-\theta_B$, where $\theta_{X}$ are the controlled laser phase offsets at the two clocks.  If we consider only atomic projection noise in $p_A$ and $p_B$, this measurement has a variance of $var(\delta\phi_{AB})=\frac{1}{N_A}+\frac{1}{N_B}$, where $N_A$ and $N_B$ are the numbers of atoms in clocks $A$ and $B$. The fractional frequency stability of the clock comparison is then $\sigma_y(\tau)=\sqrt{var(\delta\phi_{AB})}/(2\pi\nu \sqrt{T\tau})$.

A complication is introduced by the fact that $\phi_L$ will be initially unknown, which leads to ambiguities in the trigonometric inversions from which $\delta\phi_{AB}$ is calculated.  Such ambiguities will be absent in the majority of measurements, if an approximate value of $\delta\phi_{AB}$ can be determined through prior calibrations (with $var(\delta\phi_{AB})\ll 1)$, and phase offsets $\theta_{X}$ are adjusted such that $\phi_A-\phi_B-(\theta_A-\theta_B)\approx\pi/2$.  After this calibration procedure, $p_A$ and $p_B$ represent approximate quadratures of the laser-atom phase difference, and for most values of $\phi_L$ the trigonometric inversions are unambiguous. In such a measurement the Ramsey free-evolution time is no longer constrained by laser decoherence, and the Dick effect due to the probe source is absent.  Therefore, more rapid frequency comparisons of similar-frequency many-atom optical clocks should also be possible.

Small values of $\sigma(\tau)$ in frequency comparisons are useful for evaluating and improving the performance of optical clocks and for metrological applications.  For example, comparison of clocks in geographically distinct locations can be used to evaluate spatial and temporal variations in the geoid \cite{Chou2010Relativity}. More generally, any physical process that leads to small, constant frequency shifts in an optical clock can be studied in this way.  This includes relativistic effects as well as shifts caused by electric fields, magnetic fields and atom collisions. Our observation of a Q-factor beyond $10^{15}$ and a frequency ratio measurement instability of $3.7\times10^{-16}/\sqrt{\tau/s}$ highlights the intrinsic sensitivity of optical clocks as a metrological tool.

This work is supported by ONR, AFOSR, DARPA, and IARPA. We thank D. Leibrandt and J. Sherman for comments on the manuscript. Publication of NIST, not subject to U.S. copyright.

\end{document}